\documentclass[preprint,showpacs,preprintnumbers,amsmath,amssymb,12 pt.,one-column,double-spaced]{revtex4-1}

\usepackage{graphicx}
\usepackage{dcolumn}
\usepackage{bm}
\usepackage{color}

\begin{document}

\title{Controllable structuring of exciton-polariton condensates in cylindrical pillar microcavities}

\author{V.K.~Kalevich$^{1,2}$}

\author{M.M.~Afanasiev$^{1,2}$}

\author{V.A.~Lukoshkin$^{1,2}$}

\author{D.D.~Solnyshkov$^3$}

\author{G.~Malpuech$^3$}

\author{K.V.~Kavokin$^{1,2}$}

\author{S.I.~Tsintzos$^4$}

\author{Z.~Hatzopoulos$^5$}

\author{P.G.~Savvidis$^{4,6}$}

\author{A.V.~Kavokin$^{1,7}$}

\affiliation{$^1$Spin Optics Laboratory, State University of Saint-Petersburg, 1\, Ulianovskaya, 198504 St-Petersburg, Russia}

\affiliation{$^2$Ioffe Institute, Russian Academy of Sciences, 26\,Politechnicheskaya, 194021 St-Petersburg, Russia}

\affiliation{$^3$Insitut Pascal, Photon-N2, Clermont
Universit\'{e}, CNRS and University Blaise Pascal, 24 avenue des
Landais, 63177 Aubiere cedex, France}

\affiliation{$^4$IESL-FORTH, P.O. Box 1527, 71110 Heraklion,
Crete, Greece}

\affiliation{$^5$Department of Physics, University of Crete, 71003
Heraklion, Crete, Greece}

\affiliation{$^6$Department of Materials Science and Technology,
University of Crete, 71003 Heraklion, Crete, Greece}

\affiliation{$^7$Physics and Astronomy School, University of
Southampton, Highfield, Southampton SO171BJ, United Kingdom}

\begin{abstract}
We observe condensation of exciton polaritons in quantum states
composed of concentric rings when exciting cylindrical pillar
GaAs/AlGaAs microcavities non-resonantly by a focused laser beam
normally incident at the center of the pillar. The number of rings
depends on the pumping intensity and the pillar size, and may
achieve 5 in the pillar of 40\,$\mu$m diameter. Breaking the axial
symmetry when moving the excitation spot away from the pillar
center leads to transformation of the rings into a number of
bright lobes corresponding to quantum states with nonzero angular
momenta. The number of lobes, their shape and location are
dependent on the spot position. We describe the out-of-equilibrium
condensation of polaritons in the states with different principal
quantum numbers and angular momenta with a formalism based on
Boltzmann-Gross-Pitaevskii equations accounting for repulsion of
polaritons from the exciton reservoir formed at the excitation
spot and their spatial confinement by the pillar boundary.
\end{abstract}



\maketitle

\section{Introduction.}
Exciton-polaritons are mixed light-matter quasiparticles that
appear due to exciton-photon coupling in semiconductor crystal
structures \cite{Agran, Hopf}. They are composite bosons and
demonstrate characteristic bosonic effects, in particular,
stimulated scattering \cite{Savvidis2000} and Bose-Einstein
condensation \cite{KasperNature2006, Balili2007}. These effects
are at the origin of polariton lasing which is manifested by
spontaneous generation of coherent and monochromatic light by a
many-body coherent state: condensate of exciton-polaritons
\cite{Imam1996, Christopoul2007}. Polariton lasers are promising
for applications in opto-electronics and information technologies
\cite{AVNaturePhot2013}. From the point of view of fundamental
physics, they represent a unique laboratory for studies of
coherent many-body quantum systems. The microscopic size of
polariton condensates (typically, $0.01 \div 0.1$\,mm) allows
studying their shapes by optical spectroscopy methods. Another
important aspect of polariton condensation is that it does not
necessarily take place in the ground state. Very clear evidences
of polariton condensation in excited states were given in Refs.
\cite{Bajoni_PRL, Wertz2010, Wertz2012}. The non-resonant optical
pumping creates an excitonic reservoir localized under the
excitation spot. In the mean-field approximation, the interactions
of excitons from the reservoir and exciton-polaritons from the
condensate may be accounted for by introducing an effective
repulsive potential acting upon the polaritons \cite{Wertz2010,
Wouters2008, Christmann2012}, which pushes polaritons away from
the pump spot. Exploiting this effect, in the recent years,
several groups reported shaping of polariton condensates in
optical traps \cite{Tosi2012, Askit2013}, strain induced traps
\cite{Liu_arx}, laterally patterned microcavity structures
realized by etching
\cite{Wertz2010,Sala_arx,Ferrier2011,Galbiati2012}, or metal
deposition \cite{Lai2007,Bruckner2012}. A pattern formation in
exciton-polariton condensates has been observed by several groups
\cite{Manni2011, Ferrier2011, Dreismann2014}. In particular,
formation of ring condensates with large angular momenta has been
demonstrated in etched rings microcavities\cite{Sturm2014} and
using optically created confining potentials \cite{Dreismann2014}.
The same type of optical potential has been used to create
polariton condensates with large angular momenta which led to the
formation of the ``Abrikosov like" vortex chain
\cite{Boulier_arx}. Ring-shape condensates of exciton-polaritons
attract a specific interest as they are promising for generation
of non-dissipative Bessel-Gaussian light beams \cite{Ryu2014} and
may be used for Mach-Zehnder, Sagnac \cite{Sturm2014} or
Aharonov-Bohm type interferometry \cite{Shelykh2009} and for
studies of the vortex lattice dynamics \cite{Boulier_arx}.


Recently, we have reported an observation of ring-shape
condensates in cylindrical pillar microcavities
\cite{Kalevich2014}. The rings were formed due to repulsion of the
polaritons from the excitonic reservoir created in the central
part of the pillar by non-resonant optical excitation. Together
with the quantum confinement potential provided by the pillar
boundary, this optically induced repulsive potential governs the
shape of polariton condensates. The eigenmodes in the cylindrical
potential of an empty pillar are typically characterized by a
radial quantum number which sets the number of rings visible in an
experiment, and an angular momentum quantum number setting the
number of lobes which can appear on the rings. Of course, the
exact shape of the condensate wave function depends on the shape
of the effective potential induced by the exciton reservoir, on
its position in the real space and on the strength of the
interaction between polaritons. The radial and angular-momentum
quantum numbers can be used to characterize the polariton
condensate state in the majority of cases. Here we show that,
tuning the diameter of the pillar, intensity and position of the
pumping beam, one can achieve a remarkable degree of control over
these quantum numbers, which fully govern the shape of polariton
condensates. In particular, we demonstrate the formation of
condensates composed by multiple concentric rings where the
excitation spot is at the center of the pillar. The number of
rings is set by the diameter of the pillar and the pumping
intensity and may be varied between 1 and 5 in our experiments.
Furthermore, the formation of condensates in a superposition of
states with large angular momentum magnitude, i.e. the absolute
value of its projection on the structure axis, has been achieved
by shifting the excitation spot from the center of the pillar. The
condensate in this case splits into a number of lobes which are
symmetrically distributed on a ring. When the pumping spot is
strongly shifted away from the pillar center, the cylindrical
symmetry of the potential seen by the condensate is fully broken
and the condensate forms a pattern reminding a ``heart" pictogram.
These drastic modifications of the shape of the wave-functions of
polariton condensates are fully controllable. We have modelled the
system with Boltzmann-Gross-Pitaevskii equations
\cite{Galbiati2012} which takes into account pumping, decay,
energy relaxation and the spatial profile of the condensate wave
function in the potential created by the pillar, the cloud of
non-condensed excitons and the condensate itself. A good agreement
between theory and experiment has been obtained.

This paper is organized as follows. In Sec.~II we describe the
sample and the experimental setup. The experimental results are
presented in Sec.~III, where we demonstrate a dramatic dependence
of the polariton-condensate pattern on the diameter of the pillar,
pumping power and the position of the excitation spot. The
theoretical model, results of calculation and their comparison
with the experiment are presented in Sec.~IV. The summary of our
findings is given in Sec.~V.

\section{SAMPLE AND SETUP}
We studied a set of cylindrical pillars which were etched from a
planar 5$\lambda$/2 AlGaAs distributed Bragg reflector
microcavity. The measured quality factor is $Q=16000$. Four sets
of three 10\,nm GaAs quantum wells (QWs) are placed at the
antinodes of the cavity electric field to maximize the
exciton-photon coupling \cite{Tsotsis2012}. The microcavity wedge
allowed scanning across the sample to set the detuning energy
$\delta=E_{C}-E_{X}$, where $E_{C}$ and $E_{X}$ are energies of
the cavity mode and of the heavy-hole exciton at zero in-plane
wavevector ($k = 0$). The photoluminescence (PL) of the pillar was
non-resonantly excited by a cw Ti:sapphire laser tuned to the
local minimum of the upper stop-band of the distributed Bragg
reflector ($\approx 110$\,meV above $E_{X}$). The laser beam was
focused to a 2\,$\mu$m spot by a microscope objective (numerical
aperture $=0.42$). The same objective was used to collect the PL
signal. Real space images as well as the \emph{k}-space images of
pillars were projected on the entrance 100\,$\mu$m slit of a 50
cm-monochromator and, after spectral dispersion, were recorded by
a cooled CCD-camera. When taking real space and \emph{k}-images
without spectral dispersion, the grating of the monochromator was
set to the zeroth order of diffraction, and the width of the
entrance slit was set to 3\,mm. All experiments were performed at
normal incidence of the excitation beam on the sample. A cut-off
interference filter was installed in front of the entrance slit to
suppress the excitation laser light scattered from the pillar
surface. The sample under study was kept in the helium-flow
cryostat at $T=3.5$\,K.

We studied pillars with the diameters of 25, 30, and 40\,$\mu$m.
The pillars chosen for this study are characterized by a negative
photon-exciton detuning $\delta = - (0.5 \div 3.5)$\,meV. We have
found no strong qualitative dependence of the observed effects on
the detuning in the limited range of detunings we have examined.

\section{EXPERIMENTAL RESULTS}
\subsection{Splitting of polariton condensate into concentric rings }

Figure 1 presents real-space images of the (a) 25, (b) 30 and (c)
40\,$\mu$m pillars obtained under sharply focused nonresonant
excitation at the center of the pillar using a pumping power 1.5
times higher than the polariton condensation threshold (an example
of experimental determination of the condensation threshold
$P_\textrm{th}$ is given below, see the discussion of Fig.~9).
Every image represents a set of concentric rings, the number of
which increases with the increase of the pillar diameter. The
inner diameter of the outer ring in every pillar coincides with
that of the pillar top shown by the dashed circumference. In
Figures~1a and 1b the outer ring emission is much weaker than the
emission of the inner rings. The spectral analysis of the
real-space images in Fig.~1, similar to one described in detail in
the Ref. \cite{Kalevich2014} for the 25\,$\mu$m pillar, shows that
the outer ring is due to the emission of heavy-hole excitons,
which escapes from the side surface of the pillar. Thus, the outer
ring is not related to the condensate, so that the number of
condensate rings seen in the Fig.~1 amounts to 1, 2 and 5 for the
pillars of the diameter of 25, 30 and 40\,$\mu$m, respectively.

\begin{figure*}[t]
    \centering
    \includegraphics[width=0.8\linewidth]{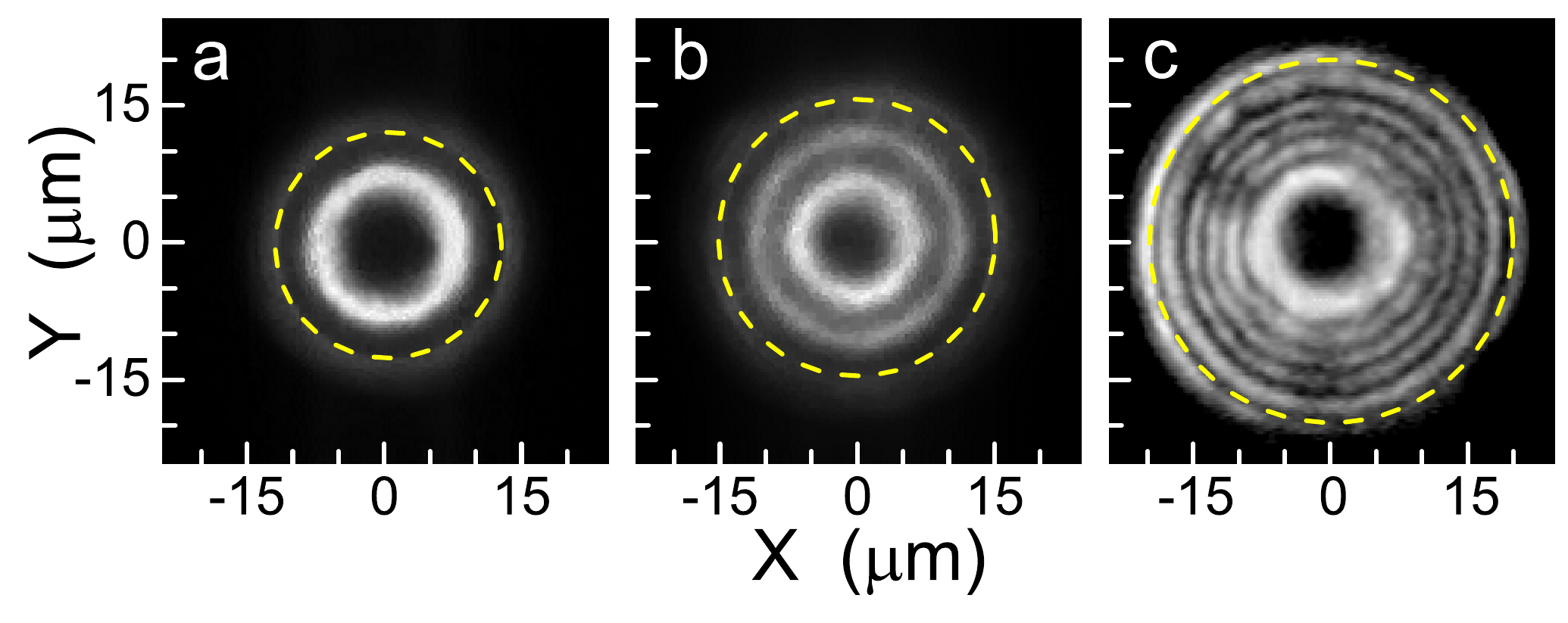}
    \caption{  Real-space images of (a) 25, (b) 30 and (c) 40\,$\mu$m pillars under nonresonant excitation
    into the center of the pillar. The excitation energy $h\nu_{\textrm{exc}}=1.664$\,eV, pumping
    power $P \approx 1.5P_\textrm{th}$, $T=3.5$\,K. Dashed circumferences show pillar edges.}
    \label{Fig1}
\end{figure*}

\begin{figure*}[t]
    \centering
        \includegraphics[width=0.7\linewidth]{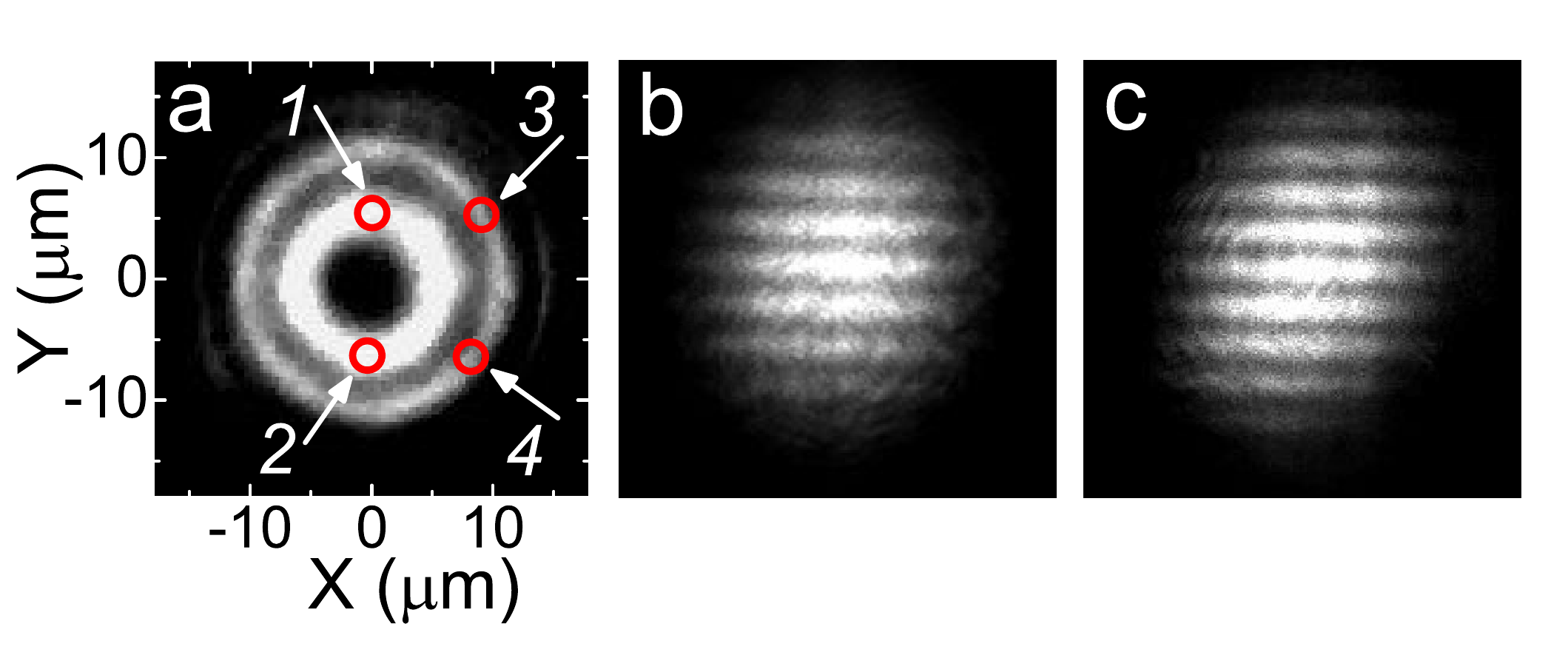}
    \caption{(Color online) (a) Real-space image of 30\,$\mu$m pillar under optical excitation into
    the center of the pillar and the Young interferometry patterns obtained by superimposing the light beams
    emitted by spots (b) \textit{1}--\textit{2} and (c) \textit{3}--\textit{4}. $P/P_\textrm{th}\approx2$. }
    \label{Fig2}
\end{figure*}

In order to make sure that each of the inner rings belongs to a
single condensate, we have performed the Young interferometry
measurements, as described in Ref. \cite{Kalevich2014}. The
coherence in each of the two inner rings in the 30\,$\mu$m pillar
has been checked by measuring the interference pattern from the
spots marked by small circles \textit{1}--\textit{4} in Fig.~2a.
The images of spots \emph{1} and \emph{2} for the smaller ring and
of spots \textit{3} and \textit{4} for the larger ring have been
superimposed on the detector using a supplementary lens. The
corresponding fringe patterns are clearly seen in the
interferometry images shown in Figs.~2b and 2c. This observation
confirms the buildup of spatial coherence in every ring of the
fragmented polariton condensate.

\begin{figure*}[b]
    \centering
    \includegraphics[width=0.9\linewidth]{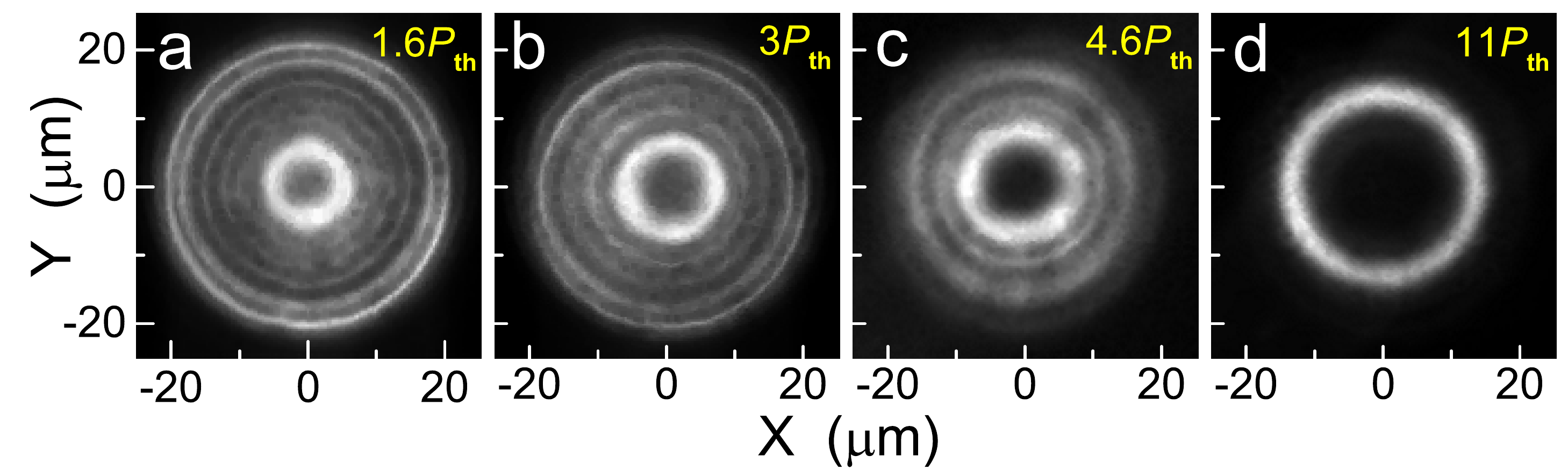}
    \caption{Transformation of 5 rings into one with the pumping power increase. The micro-PL images are obtained
    under excitation into the center of 40\,$\mu$m pillar.
    $P/P_\textrm{th}=1.6$ (a), 3 (b), 4.6 (c) and 11 (d).  } \label{Fig3}
\end{figure*}

The number of rings we observe in micro-PL has been found to
depend on the diameter of the pillar and the pumping power. As a
general rule, the higher the pumping power the lower is the number
of rings. This can be seen in Fig.~3, which demonstrates the
reduction of the number of the rings in emission of the 40\,$\mu$m
pillar from 5 to 1, as the pump power increases from
$1.6P_\textrm{th}$ to $11P_\textrm{th}$. One can see that the ring
of the smallest diameter has the highest brightness (Figs.~3a--c).
As the pumping intensity increases, the diameter of this ring
increases as well (approximately, from 10 to 25\,$\mu$m), and
simultaneously the rings of larger diameters disappear one by one.
This is a result of the increase both of the height of the
potential induced by the cloud of uncondensed excitons and of the
polariton relaxation rate. Indeed, close to the threshold the
condensation takes place in excited states, while far above
threshold it takes place in the ground state of the potential trap
formed by the pump and the pillar boundary. The evolution of
polariton condensates towards the ground state with the pump
intensity increase is a general feature reported in several
previous publications \cite{Wertz2010, Wertz2012, Sala_arx,
Galbiati2012}.

\begin{figure*}[b]
    \centering
    \includegraphics[width=0.9\linewidth]{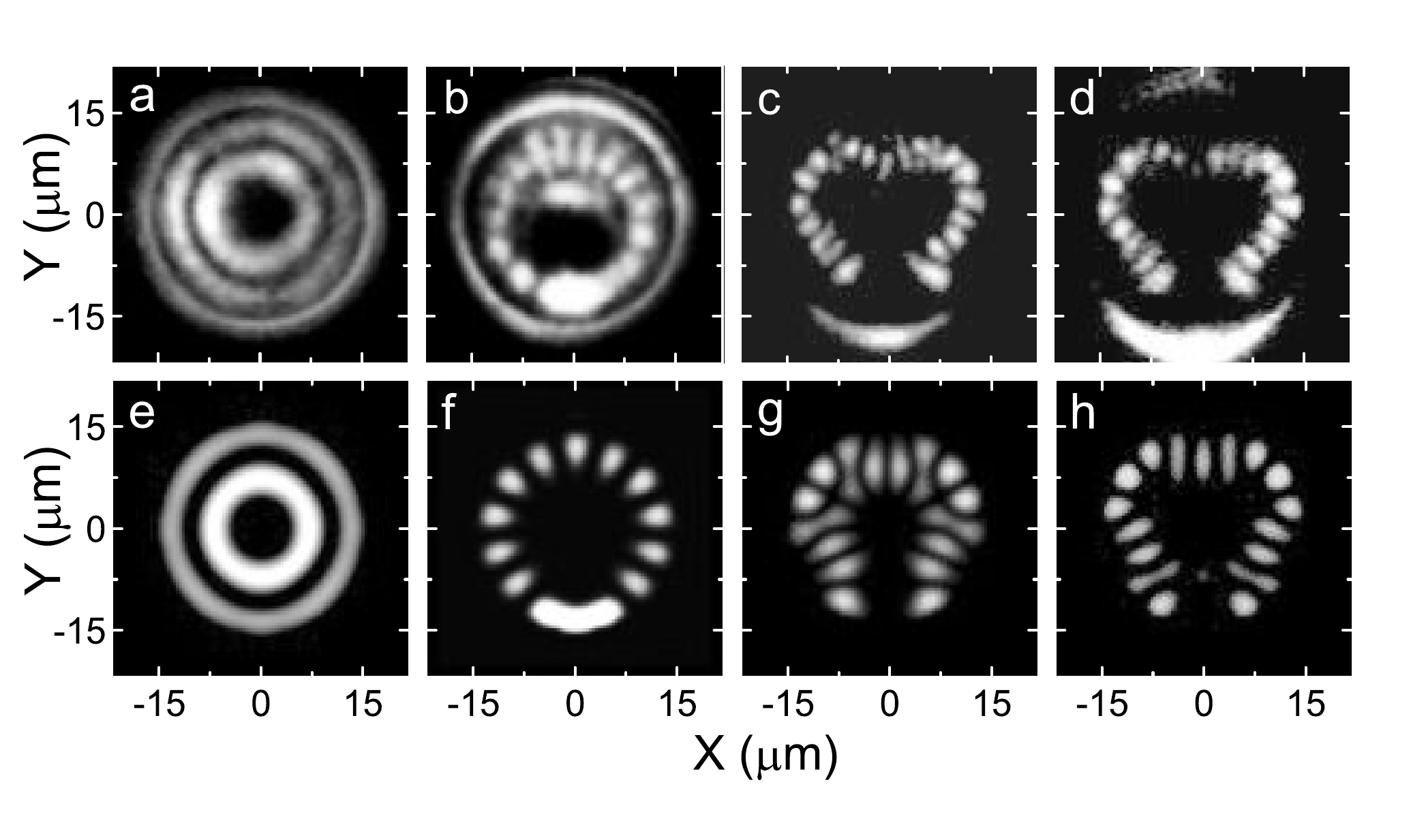}
    \caption{Experimental (a--d) and calculated (e--h) real-space images of 30\,$\mu$m pillar
    at different shifts of the excitation spot downwards from the pillar center. The shift is equal to 0 (a, e), 3\,$\mu$m (b, f),
    9\,$\mu$m (c, g), and 13\,$\mu$m (d, h). $P/P_\textrm{th}=1.7$. } \label{Fig4}
\end{figure*}

\subsection{Azimuthal structuring of the condensate}

Once we shift the pump spot slightly away from the center of the
pillar, the axial symmetry of the potential seen by the polariton
condensates appears to be broken. As a consequence, the condensate
rings transform into periodic sequences of bright lobes as can be
seen in Figs.~4 and 5 for the pillars of 30 and 40\,$\mu$m
diameters, respectively.

Figure 4 shows spatially resolved PL images of the 30\,$\mu$m
pillar excited slightly above the polariton condensation threshold
for different pump spot positions. One can see that the two
concentric rings, observed when exciting at the center of the
pillar (Fig.~4a), transform into a set of well-resolved lobes
(Fig.~4b--d). If the shift of the excitation spot from the center
of the pillar is small (3\,$\mu$m), these lobes still form a ring
(Fig.~4b), while if the excitation spot is shifted further towards
the pillar edge, they form a complex pattern (Figs.~4c, d), which
reminds a pictogram of heart.

\begin{figure*}[t]
    \centering
    \includegraphics[width=0.9\linewidth]{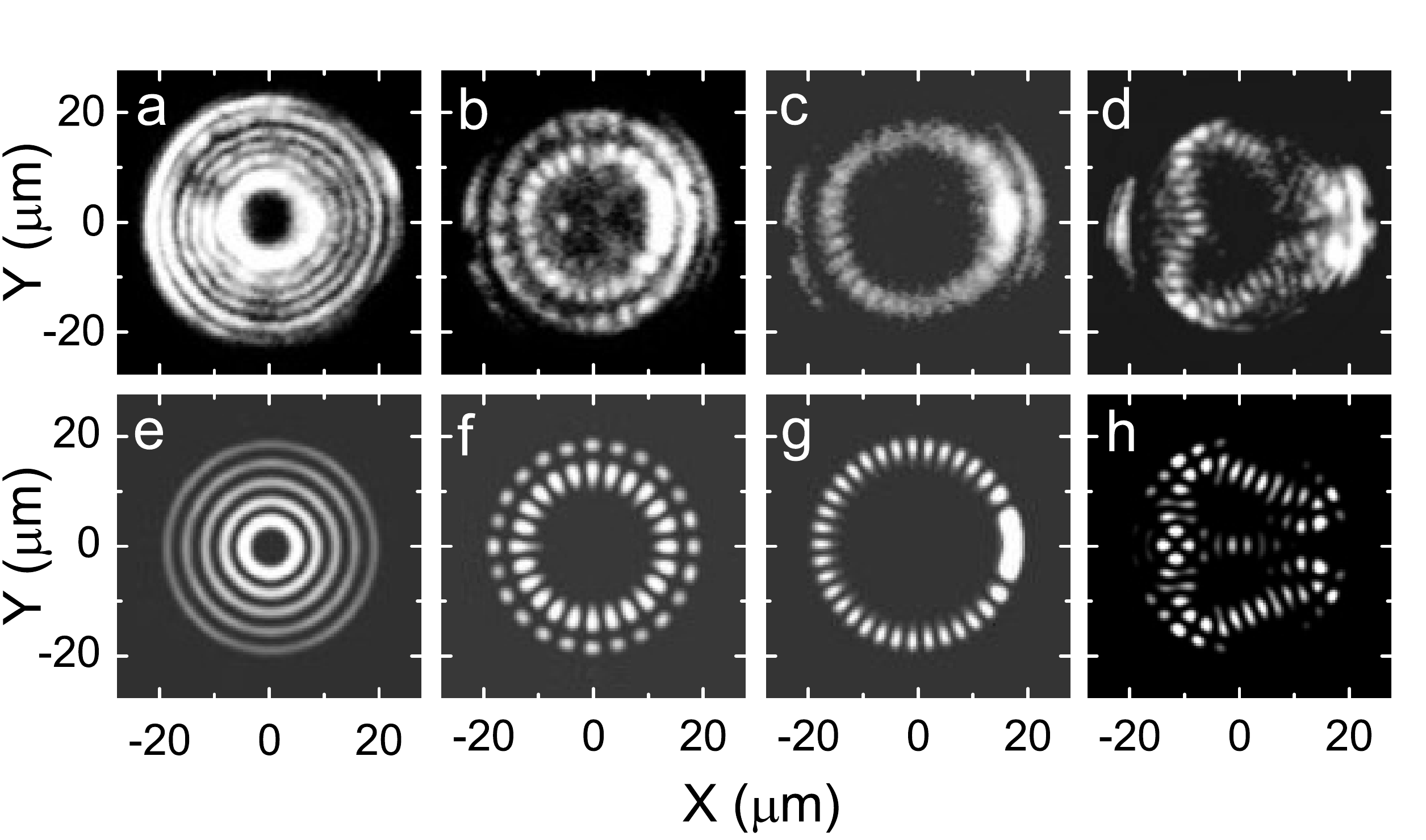}
    \caption{Experimental (a--d) and calculated (e-–h) real-space images of the polariton condensates in the 40\,$\mu$m pillar
    at different shifts of the excitation spot to the right from the pillar center. The shift is equal to 0 (a, e), 5\,$\mu$m (b, f),
    10\,$\mu$m (c, g), and 17\,$\mu$m (d, h). $P/P_\textrm{th}\approx1.5$. } \label{Fig5}
\end{figure*}

\begin{figure}[t]
    \centering
    \includegraphics[width=0.6\linewidth]{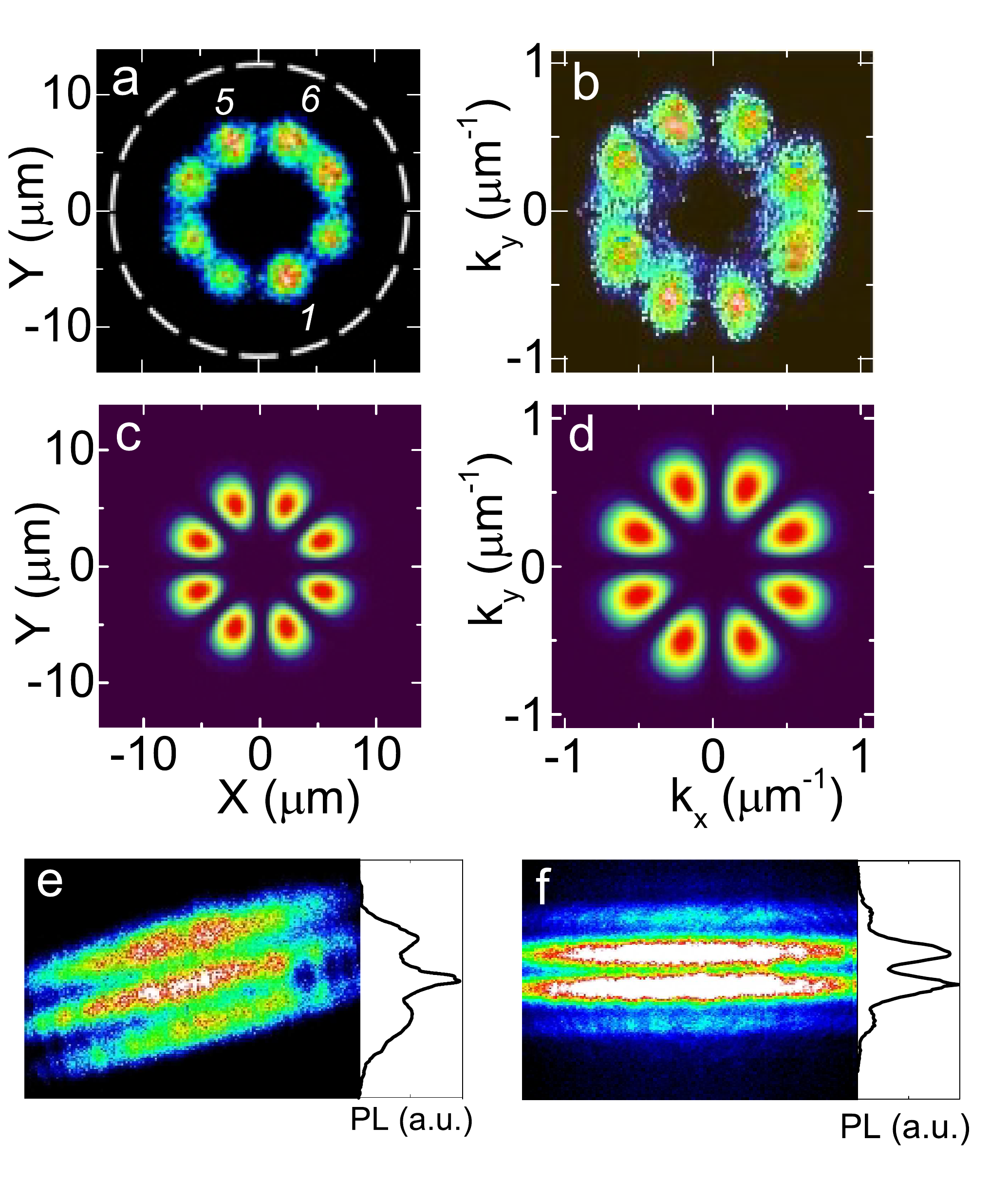}
    \caption{(Color online) Splitting of the polariton condensate into 8 lobes observed
    in (a) real-space and (b) \emph{k}-space images of a 25\,$\mu$m pillar at the shift of $\sim$\,1\,$\mu$m,
    $P/P_\textrm{th} \approx 1.7$. (c) and (d) are calculated real-space and \emph{k}-space images.
    (e, f) are Young interferometry patterns obtained by superimposing light emitted by lobes (e) \textit{1}
    and \textit{5} and (f) \textit{1} and \textit{6}. Right insets in figures (e) and (f) present PL intensity profiles
    at the centers of the interference images (e) and (f).} \label{Fig6}
\end{figure}

\begin{figure*}
    \centering
    \includegraphics[width=0.9\linewidth]{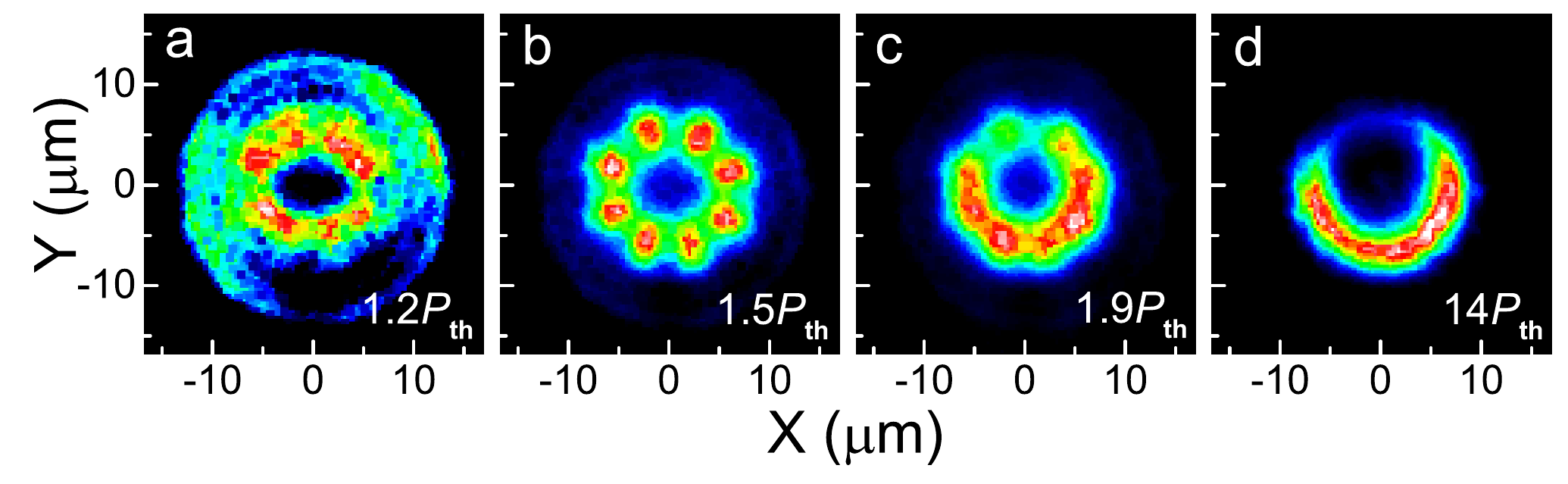}
    \caption{(Color online) Transformation of 8 lobes into a crescent with the increase of the excitation power
    in a 25\,$\mu$m pillar. $P/P_{\textrm{th}}$ is equal to 1.2\,(a), 1.5\,(b), 1.9\,(c) and 14\,(d).} \label{Fig7}
\end{figure*}

Even more complex lobe patterns have been observed in a 40\,$\mu$m
pillar, as Fig.~5 shows. Five uniform rings, observed when
exciting in the center of the pillar (Fig.~5a), transform into two
rings of lobes (Fig.~5b), which merge to form a single ring of
lobes (Fig.~5c) as the pump spot shifts by 5 and 10\,$\mu$m from
the center of the pillar, respectively. The further shift of the
pump spot towards the pillar edge (17\,$\mu$m shift in Fig.~5d)
results in the redistribution of the lobes which now form the
picture of a fly with its head oriented towards the pump spot.

It is important to check the spatial coherence of the condensate
split into many lobes. We have performed the Young interferometry
measurements in a 25\,$\mu$m pillar, where 8 clearly
distinguishable lobes separated by sufficiently large distances
have been observed (Figs.~6a and 6b). Simulations of the polariton
and exciton dispersions in the model of three coupled oscillators
(not shown here) yield the detuning value of $\delta =-3$\,meV in
this particular pillar. The pillar with a smaller diameter has
been chosen for these measurements as it shows larger lobes
separated by larger distances as compared to 30 and 40\,$\mu$m
pillars. The experimental real-space images as well as
\emph{k}-space images, taken in the 25\,$\mu$m pillar for a very
small (not exceeding 1\,$\mu$m) displacement of the pump spot from
the center, show 8 nearly identical lobes distributed
symmetrically on one ring. Note that the maxima of emission of
each of the 8 lobes in the \emph{k}-space image in Fig.~6b
correspond to the observation angle of about 6 degrees (for
details see below the discussion of Figs.~7 and 8). Thus, the
emission of the lobe condensate in the far field range is composed
by 8 beams, oriented at small ($\sim\,6^\textrm{o}$) oblique
angles with respect to the axis of the structure.

The Young interferometry has been realized by combining the
magnified real space images of two lobes passing through two
50\,$\mu$m holes in a screen placed between the sample and the
detector. In panels (e) and (f) in Fig.~6 one can see the Young
interferometry patterns obtained by superimposing light beams
emitted by lobes \textit{1} and \textit{5} and \textit{1} and
\textit{6}, respectively. Very clear interference patterns seen in
these images confirm the spatial coherence of the observed lobe
condensate. Right insets in panels (e) and (f) present PL
intensity profiles at the centers of the interference images (e)
and (f). The intensity profile in the inset (e) has a maximum in
its center, while the similar profile in the inset (f) has a
minimum in its center. Clearly, the interference of the lobes
\textit{1} and \textit{5} is constructive, while the interference
of the lobes \textit{1} and \textit{6} is destructive. This means
that the lobes \textit{1} and \textit{5} are in phase, while the
lobes \textit{1} and \textit{6} are in anti-phase. We have checked
in a similar way (not shown here), that the lobes \emph{1} and
\emph{4} have opposite phases as well. This proves that the
nearest neighboring lobes are in the anti-phase. Having in mind
the even number of lobes in Fig.~6a, one can conclude that the
wave-function of the condensate on a ring is nothing but a
standing wave resulting from the interference between two waves
rotating in clockwise and counterclockwise directions, in
agreement with Ref. \cite{Dreismann2014}.

The lobe condensate pattern is very sensitive to the excitation
power as one can see in Fig.~7 presenting the real space images of
polariton lasing from the 25\,$\mu$m pillar taken at different
pump powers. The formation of the lobes is starting from the pump
powers just above the polariton condensation threshold (Fig.~7a).
The further increase of pumping intensity results in a formation
of a very clear and symmetric image constituted by 8 equidistant
and almost identical lobes (Fig.~7b). The further pumping power
increase makes lobes to merge (Fig.~7c) and, eventually, to form a
crescent pattern (Fig.~7d).

To obtain the energy spectrum of the above mentioned condensate,
we have measured the energy dependence of the angular distribution
of the polariton emission as a function of the pumping power for
the position of the pump spot at the surface of the pillar
corresponding to the formation of a condensate containing 10
lobes. We have focused the \emph{k}-space image of the condensate
to the entrance 100\,$\mu$m slit of the monochromator, in such a
way that the slit passed through the centers of two lobes situated
most closely to the vertical axis crossing the center of the
condensate image in the \emph{k}-space. The width of the angular
distribution of emission measured in this way is different by no
more than by 5\% from one that would have been measured along the
diameter.

The polariton dispersion curves, measured in this way, are shown
in Fig.~8. At the pumping power below the condensation threshold,
the polariton dispersion exhibits a minimum at 1.536\,eV and $k=0$
(Fig.~8a). A broad emission line at 1.543\,eV is interpreted as
the PL of heavy-hole excitons coming from the side surface of the
pillar due to the Rayleigh scattering of light at the
imperfections of this surface, as we have already found
\cite{Kalevich2014} when exciting into the pillar center. This
edge emission is responsible for the outer rings observed in
Fig.~1.

\begin{figure}
        \centering
        \includegraphics[width=0.5\linewidth]{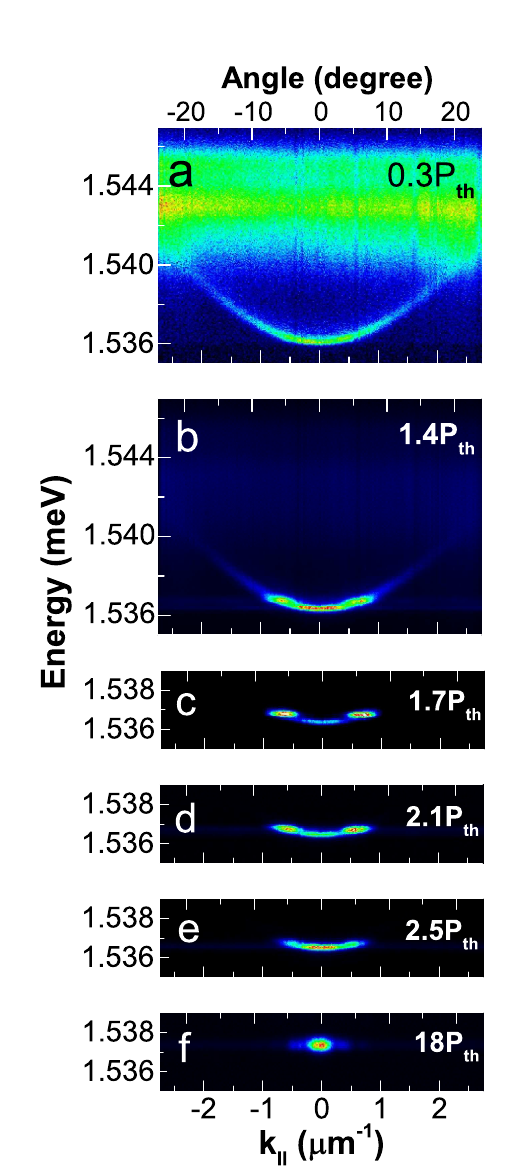}
        \caption{(Color online) Exciton-polariton dispersion curves in the 25\,$\mu$m pillar taken at
        different excitation powers and at the shift of the excitation spot from pillar center corresponding
        to the observed 10 lobes. $h\nu_{\textrm{exc}}=1.662$\,eV, $T=3.5$\,K, $\delta=-3$\,meV.
        The brightness in each image is normalized to its maximum.} \label{Fig8}
\end{figure}

The increase of the excitation power above the condensation
threshold results in an abrupt narrowing of the distribution of
polariton emission as a function of energy and angle, down to
fractions of meV and a couple of angular degrees as can be seen in
Figures 8\,b--f. This effect is a fingerprint of polariton
condensation. We remind that, if the excitaton spot is placed at
the center of this pillar, the condensation in a single ring takes
place. The corresponding reciprocal space emission
\cite{Kalevich2014} is localized around $k = 0$. When the
excitation spot is slightly moved away, as we show here, the
emission just above threshold is composed by the peak at $k=0$ and
two supplementary maxima, at the detection angles of
$6.5^\textrm{o}$, and the energy exceeding the bottom of the low
polariton branch by about 0.3\,meV (Fig.~8b). The intensity of the
supplementary peaks in Figure 8 rapidly increases with the
increase of the pumping power, so that they dominate the central
peak at $P = 1.7P_\textrm{th}$ and $P = 2.1P_\textrm{th}$, as one
can see in the Fig.~8c and Fig.~8d, respectively. This observation
is correlated with the real space images shown in Figure~7. Just
above threshold, the lobe structure is partly hindered by the
presence of other energy states, whereas it fully dominates around
$1.5P_\textrm{th}$. At $P=2P_\textrm{th}$ the supplementary maxima
start approaching each other (Fig.~8d) and they merge eventually
at $P = 2.5P_\textrm{th}$ (Fig.~8e). The further pumping increase
until $P = 18P_\textrm{th}$ leads to the decrease of the angular
width of emission down to $\approx 2.5^\textrm{o}$ (Fig.~8f),
whereas in real space, the lobes disappear. These data allow
visualizing the passage from out-of-equilibrium exciton-polariton
condensation in the excited states just above threshold to the
condensation in the ground energy state of the potential imposed
by the exciton cloud and the pillar boundary.

\begin{figure}
        \centering
        \includegraphics[width=0.5\linewidth]{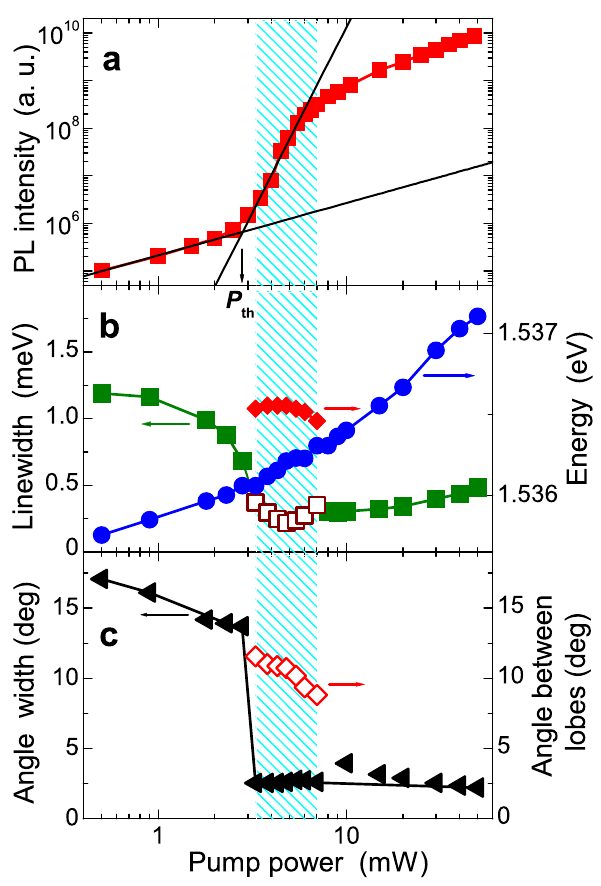}
        \caption{(Color online) Dependences of (a) integrated intensity, (b) energy and spectral width (squares)
        and (c) angular width of polariton PL (triangles) and angle between the angle-resolved emission maxima
        corresponding to two opposite lobes (diamonds) on
        the pump power in the 25\,$\mu$m pillar at such a shift ($\sim 1\,\mu$m) of excitation spot from
        the pillar center that 10 lobes are observed. Diamonds in panel (b) show PL peak energies of the lobes,
        while circles present polariton ground state energy. Vertical arrow in panel (a) shows the threshold
        of polariton stimulated emission $P_\textrm{th}=2.8$\,mW. Solid lines are guides for the eye.} \label{Fig9}
\end{figure}

Figure 9 presents the detailed power dependences of the main
parameters extracted from the polariton PL spectra measured in the
25\,$\mu$m pillar at the same shift of the excitation spot as in
Fig~8. The dependence of the integrated PL intensity on the
pumping intensity is shown in Fig.~9a. It demonstrates a very
strong nonlinear increase. The PL intensity rises by two orders of
magnitude when the excitation power is only doubled, from 3 to
6\,mW. The threshold behavior of the PL intensity is a signature
of the polariton condensation regime. The threshold power found
from this curve is equal to $P_\textrm{th}=2.8$\,mW (shown by the
vertical arrow in panel a).

Figure 9b shows the energies of the polariton ground state ($k=0$,
circles) and of the lobe condensate (diamonds). It is seen that
the lobe condensate exists within the narrow range of pumping
powers, approximately from 3 to 7~mW, which is hatched in Fig.~9.
Its energy is nearly constant throughout this power range. At the
same time, the ground state energy of exciton-polaritons
monotonously increases, reproducing the well-known blue shift
induced by the exciton-exciton interaction
\cite{Bajoni_PRL,Bajoni_PRB,Vladimirova}. Note that the range of
existence of the lobe condensate largely coincides with the range
of the superlinear growth of the PL intensity in Fig.~9a.

Spectral widths (FWHM) of the emission of a single lobe (open
squares) and of the integral emission of the polariton gas outside
the condensate (full squares) are also shown in Fig.~9b. The
spectral width of the lobe condensate varies slowly with the pump
power  and remains close to 0.3\,meV. On the contrary, the
emission of the non-condensed polariton gas is characterized by a
much larger (1.2\,meV) spectral width. The remarkable (four-fold)
narrowing of the radiation spectrum when approaching
$P_\textrm{th}$ also evidences the realization of the polariton
condensation regime at $P > P_\textrm{th}$.

The full width at half maximum (FWHM) of the angular-resolved
polariton emission as well as the angle between the emission
maxima of two separate lobes in \emph{k}-space are shown in
Fig.~9c by triangles and diamonds, respectively. The angular width
of a single lobe is given for the range of existence of the lobe
condensate. One can see that at $P > P_\textrm{th}$ the angular
width decreases abruptly, from 13.5 to 2.5 degrees. At the same
time, the angle between the emission maxima of two separate lobes
with the opposite lateral components of the wave vector exhibits
little changes, decreasing steadily from $11.5^\textrm{o}$ to
$9^\textrm{o}$. Thus, we observe all the characteristic features
of polariton lasing: the superlinear growth of intensity and
strong narrowing of spectral and angle distributions as well as
the blue shift of the polariton emission above the lasing
threshold.

\section{THEORETICAL MODEL, RESULTS OF CALCULATION AND COMPARISON WITH THE EXPERIMENTAL DATA}
Theoretical simulations of polariton relaxation and condensation
in micropillars were carried out using the model based on coupled
semi-classical Boltzmann equations and the Gross-Pitaevskii
equation, solved in a self-consistent way. At each time step, a
stationary Gross-Pitaevskii equation is solved in order to find
the current wave functions of the polariton condensate in the
pillar, taking into account the quantum confinement due to the
edges of the pillar and the effective potential created due to the
exciton-exciton interactions between the reservoir and the
condensate as well as between polaritons in the condensate.
Neglecting spin, this equation can be written as
\begin{multline}
 E_{i,t_{j}}\Psi_{i,t_{j}}  = -
\dfrac{\hbar^{2}}{2m}\Delta\Psi_{i,t_{j}} + (\alpha \eta
\rho_{R}(x,y)N_{R}(t_{j})
 + \alpha \eta^{2}\sum_{i}|\Psi_{i,t_{j}}|^{2}n_{i,t_{j}} + V)\Psi_{i,t_{j}}
  ~.~~~~(1)\nonumber
\end{multline}
Here, $ E_{i,t_{j}} $  are  the  eigenenergies and
$\Psi_{i,t_{j}}$ are the many-body wave functions (normalized to
unity)  at the fixed time moment $t_j$, \textit{m} is the
polariton effective mass, \textit{V} is the micropillar
confinement potential, $ \alpha=6E_{b}a^{2}_{B} $ is an estimate
of the exciton-exciton interaction constant describing the
repulsion of exciton polaritons from the exciton reservoir and
from each other, $\eta $ is the exciton fraction of the considered
quantum state, $ E_{b} $ is the exciton binding energy, $a_{B}$ is
the exciton Bohr radius, $\rho_{R}(x,y)$ is the normalized
reservoir density, $N_{R}$ is the number of particles in the
reservoir and $n_{i,t_{j}}$ is the number of particles in each
quantum state.

Then, the scattering rates are calculated accounting for the
overlap between the reservoir and the wave functions found in the
previous step. The relaxation mechanisms taken into account are
both exciton-exciton and exciton-phonon scattering.
Exciton-exciton scattering is the most important for the transfer
of quasiparticles between the reservoir and the quantized
polariton states, because the energy difference in this case is
large, while exciton-phonon scattering is responsible for the
relaxation between the quantized states of the condensate, because
the energy difference between these states is small, and their
populations are relatively low as well.

As shown previously \cite{Galbiati2012}, due to the thermal
distribution of excitons in the reservoir, the exciton-exciton
interaction leads to the thermalization of the polariton states in
a similar fashion to the exciton-phonon interaction. The reservoir
is modelled as a single state with a population
$N_{R}(t_j)=n_{R}(E=0,t_j)Sm_{X}k_{B}T/2\pi\hbar^2$ (here $m_{X}$
is the exciton mass,  $S$ is the normalization surface, $n_{R}$ is
the occupation of the corresponding reservoir state), whose
thermal distribution is accounted for in the scattering rates. The
shape of the reservoir, given by the function $\rho_{R}(x,y)$, is
defined by the spatial location and shape of the pump spot (the
exciton diffusion is assumed to be slow).

The semi-classical Boltzmann equations for the quantized polariton
states matching the stationary Gross-Pitaevskii equation can be
written as:
\begin{widetext}
\begin{multline*}
~~~~\frac{{d{n_i}}}{{dt}} =  - \frac{{{n_i}}}{{{\tau _i}}} -
{n_i}\sum\limits_k {\left( {{W_{XX}}{I_{ikRR}}{N_R} +
{W_{XP}^{ik}}} \right)}({n_k} + 1)\left( {\begin{array}{*{20}{c}}
    {1,~~k < i}  \\
    {{e^{ - {E_{ik}}/{k_B}T}},~~k > i}  \\
    \end{array}} \right) \\
~~~~~~~~~~~+ ({n_i} + 1)\sum\limits_k {\left(
{{W_{XX}}{I_{ikRR}}{N_R} + {W_{XP}^{ik}}{I_{ik}}} \right)}
{n_k}\left( {\begin{array}{*{20}{c}}
    {{e^{ - {E_{ik}}/{k_B}T}},~~k < i}  \\
    {1,~~k > i}  \\
    \end{array}} \right),~~~(2)\\
+ \left( {{W_{XX}}{I_{iRRR}}{N_R} + {W_{XP}^{iR}}{I_{iR}}}
\right)\left( {{N_R}\left( {{n_i} + 1} \right) -
{n_i}\frac{{Sm_X k_{B}T}}{{2\pi {\hbar ^2}}}{e^{ - {E_{iR}}/{k_B}T}}} \right) \\
\end{multline*}
\end{widetext}
where $I_{ikRR}, I_{ik}, I_{iRRR}, I_{iR}$ are the overlap
integrals defined below, $\tau_i$ is the lifetime of the
respective state.

For the reservoir state, the equation reads:
\begin{multline}
\frac{{d{N_R}}}{{dt}} = P - \frac{{{N_R}}}{{{\tau _R}}}-
\sum\limits_i \left( {{W_{XX}}{I_{iRRR}}{N_R} +
{W_{XP}^{iR}}{I_{iR}}} \right) \times
 \left( {{N_R}\left({{n_i} + 1}
\right) - \frac{{{n_i}{S}m_X k_{B}T}}{{2\pi {\hbar ^2}}}{e^{ -
{E_{iR}}/{k_B}T}}} \right).~~(3)\nonumber
\end{multline}
Here $P$ is the non-resonant pumping populating the reservoir,
$\tau_R=400$\,ps, $W_{XX}$ is the exciton-exciton scattering rate,
which is a fitting parameter of the model, and
$W_{XP}^{ik,R}=W_{XP}\exp[-(E_{i}-E_{k,R})a_{B}/\hbar c\pi]$ is
the exciton-phonon scattering rate between the corresponding
states, $ W_{XP} $ is another fitting parameter. The details on
the derivation of these equations are given in Ref.
\cite{Galbiati2012}.

All scattering rates are affected by the corresponding overlap
integrals, which change with time together with the shape of the
quantized wave functions, according to the Gross-Pitaevskii
equation. The dimensionless overlap integrals are defined as
follows:
$I_{ikRR}=S^3\int\left|\psi_{i}^{*}\psi_{k}\right|^2\rho_{R}^2~dxdy$,
$I_{iRRR}=S^3\int\left|\psi_{i}\right|^2\rho_{R}^{3}dxdy$. The
overlap integrals obtained by summation of scattering rates with
different phonon wavevectors read:
$I_{ik}=S\int\left|\psi_{i}^{*}\psi_{k}\right|^2dxdy$,
$I_{iR}=S\int\left|\psi_{i}\right|^2\rho_{R}dxdy $. These overlap
integrals are of a significant importance, because they define
relative scattering efficiencies for different quantized polariton
states.

The results of the simulations reproducing the experimental data
are shown in the theoretical panels of Figures 4, 5, 6. Depending
on the size of the pillar, the calculated shapes of the
condensates are different, but the agreement with the experiment
is always good.

Figure 4 presents the results of theoretical simulations of the
shapes of polariton condensates compared with the experimental
data for a 30\,$\mu$m pillar (panels e--h). The transition between
the states with zero angular momentum (panel e) and high angular
momentum magnitude is obtained (panel f) by breaking of the
cylindrical symmetry due to the pump spot displacement. At large
displacements (panels g, h), the state of the condensate can be
interpreted as two opposite polariton flows generated at the
pumping spot and reflected at the pillar boundary, while the
center of the pillar remains empty because of the destructive
interference of outgoing and incoming flows.

Figure 5 presents the calculated shapes of polariton condensates
for the 40\,$\mu$m pillar (panels e--h). In the case of
centrosymmetric excitation, the condensate is formed in the
quantum state with the principal quantum number 5 (panel e). The
violation of the cylindrical symmetry leads again to the
excitation of the quantum states having higher angular momentum
magnitude (panels f, g). At the strong displacement (panel h), the
propagation of polaritons away from the pumping spot and around
the pillar is observed again, in a similar fashion to the
30\,$\mu$m pillar.

In larger pillars, the quantum confinement energy is smaller, so
that the greater number of polariton modes having different
quantum numbers can be excited simultaneously. This is why only 2
rings are observed in the 30\,$\mu$m pillar and 5 rings are
observed in the 40\,$\mu$m pillar. The polariton lasing threshold
for all these modes is achieved approximately at the same pumping
intensity. At the strong displacement, both in the 30\,$\mu$m
pillar and in the 40\,$\mu$m pillar, the theoretical simulation
shows the asymmetric pattern of lobes reminding a heart pictogram,
in a full qualitative agreement with the experiment.

Figures~6c and 6d show simulations of the exciton-polariton
condensate in the smallest, 25\,$\mu$m pillar. For this pillar, a
small displacement of the pumping spot leads to the formation of
an 8-lobe state (the principal quantum number 1, the orbital
quantum number 4). Panel (c) shows the real space image, and panel
(d) shows the reciprocal space image of this state, corresponding
to the experimental panels (a), and (b), respectively. The image
of the condensate in the reciprocal state reflects the symmetry of
the image in the real space. The absence of a maximum in the
center of the reciprocal space image clearly shows that this state
is different from the ground state, where the condensate forms at
higher pumping power, as shown in Figure~8 and reproduced in the
simulations (not shown).

The main conclusions of this theoretical section can be summarized
as follows. Because of the close polariton lasing threshold powers
for several nearly degenerate polariton modes, the condensation
may occurs not only in the thermodynamically favorable ground
state, but sometimes in the higher-energy states, whose profile is
defined mostly by the repulsive potential of the localized
reservoir. The lowest energy state is strongly repelled by the
reservoir and has a lower overlap integral with it, which is why
the scattering into this state is strongly suppressed. At the same
time, higher energy states have larger overlap integrals, but
these states experience also depletion due to the exciton-phonon
scattering towards the lower energy states. Depending on the
conditions, the condensation may start in a single state or in
several quantum states. In the latter case, at the pumping power
just exceeding the lasing threshold, the condensate occupies
several discrete energy levels. The further increase of the
pumping power leads to accumulation of particles in the lowest
energy state because of the scattering rate enhancement by bosonic
stimulation. This tendency is experimentally seen in Fig.~8.

\section{CONCLUSIONS}
We have demonstrated an efficient optical control of the shape of
exciton-polariton condensates and switching between different
topologies of the condensates in cylindrical micropillars.
Switching between multi-ring condensates and patterns of lobes is
achieved by shifting the excitation spot by fractions of a micron
from the pillar center. Such a strong sensitivity of polariton
condensates to the location of the excitation spot may be
understood in terms of switching between nearly degenerate
polariton states characterized by different radial and orbital
quantum numbers. A small perturbation is sufficient to start
condensation in a different quantum state. Bosonic condensation in
several excited states and pumping-dependent redistribution of
bosons in the real space are characteristic of driven dissipative
bosonic systems, such as polariton lasers. These results
demonstrate remarkable flexibility of wave function engineering in
polariton lasers. Multi-ring condensates have a high potentiality
for realization of Bessel-Gauss non-diffractive light beams. The
structuring of polariton condensates, their fragmentation,
formation of multiple lobes and various spatial patterns open way
to realization of bosonic intergrated circuits and logic gates.

\acknowledgements The work of VKK, MMA, VAL, KVK and AVK was
financially supported by the Russian Ministry of Education and
Science (Contract No.11.G34.31.0067). PGS acknowledges financial
support from FP7 EU ITN ``INDEX" 289968 and IRSES ``POLATER"
exchange grants and Greek GSRT ARISTEIA program Apollo. DDS and GM
acknowledge the support of ANR ``QUANDYDE" and EU ``POLAPHEN"
projects.


\begin{thebibliography}{99}

\bibitem{Agran} V.M.~Agranovich and V.L.~Ginzburg, \textit{Crystal Optics with Spatial Dispersion and
Excitons} (Springer-Verlag, New York, 1984).
%
\bibitem{Hopf} J.J.~Hopfield, Phys. Rev. \textbf{112}, 1555 (1958).
%
\bibitem{Savvidis2000} P.G.~Savvidis, J.J.~Baumberg, R.M.~Stevenson, M.S.~Skolnick, D.M.~Whittaker, and J.S.~Roberts,
Phys. Rev. Lett. \textbf{84}, 1547 (2000).
%
\bibitem{KasperNature2006} J.~Kasprzak, M.~Richard, S.~Kundermann, A.~Baas, P.~Jeambrun, J.M.J.~Keeling, F.M.~Marchetti, M.H.~Szyma\'{n}ska,
R.~Andr\'{e}, J.L.~Staehli, V.~Savona,
P.B.~Littlewood, B.~Deveaud, and Le~Si~Dang, Nature \textbf{443},
409 (2006).
%
\bibitem{Balili2007} R.~Balili, V.~Hartwell, D.~Snoke, L.~Pfeiffer, and K.~West, Science \textbf{316}, 1007 (2007).
%
\bibitem{Imam1996} A.~Imamoglu, R.J.~Ram, S.~Pau, and Y.~Yamamoto, Phys. Rev. A \textbf{ 53}, 4250 (1996).
%
\bibitem{Christopoul2007} S.~Christopoulos, G.~Baldassarri, H\"{o}ger~von~H\"{o}gersthal, A.J.D.~Grundy, P.G.~Lagoudakis, A.V.~Kavokin,
J.J.~Baumberg, G.~Christmann, R.~Butt\'{e}, E.~Feltin,
J.-F.~Carlin, and N.~Grandjean, Phys. Rev. Lett. \textbf{98},
126405 (2007).
%
\bibitem{AVNaturePhot2013} A.V.~Kavokin, Nature Photon. \textbf{7}, 591 (2013).
%
\bibitem{Bajoni_PRL} D.~Bajoni, P.~Senellart, E.~Wertz, I.~Sagnes, A.~Miard, A.~Lema\^{\i}tre, and J.~Bloch, Phys. Rev. Lett. \textbf{100}, 047401 (2008).
%
\bibitem{Wertz2010} E.~Wertz, L.~Ferrier, D.D.~Solnyshkov, R. ~Johne, D.~Sanvitto, A.~Lema\^{\i}tre, I.~Sagnes, R.~Grousson, A.V.~Kavokin,
P.~Senellart, G.~Malpuech and J.~Bloch, Nature Phys. \textbf{6},
860 (2010).
%
\bibitem{Wertz2012} E.~Wertz, A.~Amo, D.D.~Solnyshkov, L.~Ferrier, T.C.H.~Liew D.~Sanvitto, P.~Senellart, I.~Sagnes, A.~Lema\^{\i}tre, A.V.~Kavokin,
G.~Malpuech, and J.~Bloch, Phys. Rev. Lett. \textbf{109}, 216404 (2012).
%
\bibitem{Wouters2008} M.~Wouters, I.~Carusotto, and C.~Ciuti, Phys. Rev. B \textbf{77}, 115340 (2008).
%
\bibitem{Christmann2012} G.~Christmann, G.~Tosi, N.G.~Berloff, P.~Tsotsis, P.S.~Eldridge, Z.~Hatzopoulos, P.G.~Savvidis, and J.J.~Baumberg,
Phys. Rev. B \textbf{85}, 235303 (2012).
%
\bibitem{Tosi2012} G.~Tosi, G.~Christmann, N.G.~Berloff, P.~Tsotsis, T.~Gao, Z.~Hatzopoulos, P.G.~Savvidis and J.J.~Baumberg,
Nature Phys. \textbf{8}, 190 (2012).
%
\bibitem{Askit2013} A.~Askitopoulos, H.~Ohadi, A.V.~Kavokin, Z.~Hatzopoulos, P.G.~Savvidis, and P.G.~Lagoudakis,
Phys. Rev. B \textbf{88}, 041308(R) (2013).
%
\bibitem{Liu_arx} G.~Liu, D.W.~Snoke, A.~Daley, L.~Pfeiffer, and K.~West, arXiv:1402.4339.
%
\bibitem{Sala_arx} V.G.~Sala, D.D.~Solnyshkov, I.~Carusotto, T.~Jacqmin, A.~Lema\^{\i}tre, H.~Ter\c{c}as, A.~Nalitov, M.~Abbarchi, E.~Galopin,
I.~Sagnes, J.~Bloch, G.~Malpuech, and A.~Amo, arXiv:1406.4816.
%
\bibitem{Ferrier2011} L.~Ferrier, E.~Wertz, R.~Johne, D.D.~Solnyshkov, P.~Senellart, I.~Sagnes, A.~Lema\^{\i}tre, G.~Malpuech, and J.~Bloch,
Phys. Rev. Lett. \textbf{106}, 126401 (2011).
%
\bibitem{Galbiati2012} M.~Galbiati, L.~Ferrier, D.D.~Solnyshkov, D.~Tanese, E.~Wertz, A.~Amo, M.~Abbarchi, P.~Senellart, I.~Sagnes, A.~Lema\^{\i}tre,
E.~Galopin, G.~Malpuech, and J.~Bloch, Phys. Rev. Lett. \textbf{108}, 126403 (2012).
%
\bibitem{Lai2007} C.W.~Lai, N.Y.~Kim, S.~Utsunomiya, G.~Roumpos, H.~Deng, M.D.~Fraser, T.~Byrnes, P.~Recher, N.~Kumada, T.~Fujisawa, and Y.~Yamamoto,
Nature \textbf{450}, 529 (2007).
%
\bibitem{Bruckner2012} R.~Br\"{u}ckner, A.A.~Zakhidov, R.~Scholz, M.~Sudzius, S.I.~Hintschich, H.~Fr\"{o}b, V.G.~Lyssenko and K.~Leo,
Nature Photon. \textbf{6}, 322 (2012).
%
\bibitem{Manni2011} F.~Manni, K.G.~Lagoudakis, T.C.H.~Liew, R.~Andr\'{e}, and B.~Deveaud-Pl\'{e}dran, Phys. Rev. Lett. \textbf{107}, 106401 (2011).
%
\bibitem{Dreismann2014} A.~Dreismann, P.~Cristofolini, R.~Balili, G.~Christmann, F.~Pinsker, N.G.~Berloff, Z.~Hatzopoulos,
P.G.~Savvidis, and J.J. Baumberg, Proceedings of the National
Academy of Sciences of the United States of America \textbf{111},
8770 (2014).
%
\bibitem{Sturm2014} C.~Sturm, D.~Tanese, H.S.~Nguyen, H.~Flayac, E.~Galopin, A.~Lema\^{\i}tre, I.~Sagnes, D.~Solnyshkov, A.~Amo, G.~Malpuech,
and J.~Bloch, Nat. Commun. \textbf{5}, 3278 (2014).
%
\bibitem{Boulier_arx} T.~Boulier, H.~Ter\c{c}as, D.D.~Solnyshkov, Q.~Glorieux, E.~Giacobino, G.~Malpuech, and A.~Bramati, arXiv:1405.1375.
%
\bibitem{Ryu2014} C.~Ryu, K.C.~Henderson, and M.G.~Boshier, New J. Phys. \textbf{16}, 013046 (2014).
%
\bibitem{Shelykh2009} I.A.~Shelykh, G.~Pavlovic, D.D.~Solnyshkov, and G.~Malpuech, Phys. Rev. Lett. \textbf{102}, 046407 (2009).
%
\bibitem{Kalevich2014} V.K.~Kalevich, M.M.~Afanasiev, V.A.~Lukoshkin, K.V.~Kavokin, S.I.~Tsintzos, P.G.~Savvidis, and A.V.~Kavokin,
J. Appl. Phys. \textbf{115}, 094304 (2014).
%
\bibitem{Tsotsis2012} P.~Tsotsis, P.S.~Eldridge, T.~Gao, S.I.~Tsintzos, Z.~Hatzopoulos, and P.G.~Savvidis, New J. Phys. \textbf{14}, 023060 (2012).
%
\bibitem{Bajoni_PRB} D.~Bajoni, E.~Semenova, A.~Lema\^{\i}tre, S.~Bouchoule, E.~Wertz, P.~Senellart, and J.~Bloch,
Phys. Rev. B \textbf{77}, 113303 (2008).
%
\bibitem{Vladimirova} M.~Vladimirova, S.~Cronenberger, D.~Scalbert, K.V.~Kavokin, A.~Miard, A.~Lema\^{\i}tre, J.~Bloch, D.~Solnyshkov, G.~Malpuech,
and~A.~V.~Kavokin, Phys. Rev. B \textbf{82}, 075301 (2010).

\end{thebibliography}
\end{document}